\documentclass[preprint,amsmath,amssymb,11pt,english,aps,showpacs,showkeys]{revtex4}
\usepackage{graphicx}
\usepackage{graphics}
\usepackage{amsmath}
\usepackage{dcolumn}
\usepackage{amssymb}
\usepackage{bm}
\usepackage[latin1]{inputenc}

\begin{document}
\title{Wigner rotation via Fermi-Walker transport and relativistic EPR correlations in the Schwarzschild spacetime}
\author{K. Bakke}
\email{kbakke@fisica.ufpb.br}
\affiliation{Departamento de F\'isica, Universidade Federal da Para\'iba, Caixa Postal 5008, 58051-970, Jo\~ao Pessoa, PB, Brazil.}

\author{C. Furtado}
\email{furtado@fisica.ufpb.br}
\affiliation{Departamento de F\'isica, Universidade Federal da Para\'iba, Caixa Postal 5008, 58051-970, Jo\~ao Pessoa, PB, Brazil.}

\author{A. M. de M. Carvalho}
\email{ammcarvalho@gmail.com}
\affiliation{Instituto de F\'isica, Universidade Federal de Alagoas, Campus A. C. Sim\~oes - Av. Lourival Melo Mota, s/n, Tabuleiro do Martins, 57072-970, Macei\'o, AL, Brazil.}

\begin{abstract}
The Wigner rotation angle for a particle in a circular motion in the Schwarzschild spacetime is obtained via the Fermi-Walker transport of spinors. Then, by applying the WKB approximation, a possible application of the Fermi-Walker transport of spinors in relativistic EPR correlations is discussed, where it is shown that the spins of the correlated particle undergo a precession in an analogous way to that obtained by Terashima and Ueda [H. Terashima and M. Ueda, Phys. Rev. A {\bf69}, 032113 (2004)] via the application of successive infinitesimal Lorentz transformations. Moreover, from the WKB approach, it is also shown that the degree of violation of the Bell inequality depends on the Wigner rotation angle obtained via the Fermi-Walker transport. Finally, the relativistic effects from the geometry of the spacetime and the accelerated motion of the correlated particles is discussed in the nonrelativistic limit.

\end{abstract}


\maketitle

\section{Introduction}

The Wigner rotation is known in special relativity as the product of two Lorentz boots in different directions which gives rise to a boost preceded or followed by a rotation \cite{wigner,weinberg2}. Besides, the Wigner rotation is characterized by leaving the 4-momentum of the particle unchanged and making a precession of the spins in the rest frame of the particles. One effect associated with the Wigner rotation is the Thomas precession \cite{misner,thomas}. In classical physics, the Thomas precession can be viewed through a spinning object, free of any action of torque, which changes its direction in relation to that in an inertial frame due to an acceleration that stems from an external force \cite{misner}. In quantum physics, the Thomas precession corresponds to relativistic corrections in the magnitude of the spin-orbit interaction which influence in the fine structure of the atomic spectra \cite{misner,ruark,jackson}.

An interesting effect associated with the Wigner rotation is the precession of spins of the relativistic Einstein-Podolsky-Rosen (EPR) correlation \cite{epr} with respect to the initial configuration of spins due to the action of Lorentz transformations. This precession of spins yields an apparent deterioration of the initial correlations between the spins and decreases the degree of violation of the Bell inequality. As examples, it has been shown in Refs. \cite{bell1,bell2,bell3,bell4,ahn} that a decrease in the degree of the Bell inequality is yielded by the relativistic motion of the particle in the Minkowski spacetime. On the other hand, by considering a curved spacetime background, it has been shown in Refs. \cite{bell5,tu2,bcf,bcf2} that the decrease in the degree of the Bell inequality is yielded by the relativistic motion of the particles, the gravitational field and the position of the observers. Other interesting studies of quantum entanglement in curved spacetime have been made in Refs. \cite{meh,meh2,meh3,meh4,meh5,meh6}.

It is worth mentioning the work made by Borzeszkowski and Mensky \cite{repr}, where a geometric approach was proposed in order to study the relativistic EPR correlations in the presence of a gravitational field by applying the parallel transport along the world lines of the particles. However, Terashima and Ueda \cite{tu2} showed, via application of successive infinitesimal Lorentz transformations, that by taking into account the accelerated motion of the particle and the gravitational field, thus, the parallel transport cannot yield the perfect direction of the relativistic EPR correlations. From this perspective, a geometric approach based on the Fermi-Walker transport has been proposed in Ref. \cite{bcf2} in order to obtain the Wigner rotation angle and the precession of the spins of a relativistic EPR correlation. The Fermi-Walker transport was introduced in studies of quantum systems in a curved spacetime by Anandan \cite{anandan} via the WKB approximation with the aim of obtaining the geometric quantum phase which arises from the presence of torsion, the gravitational field and the Thomas precession. Other studies of geometric quantum phases associated with the Thomas precession have been made in Refs. \cite{mat1,mat2,mat3}.

The aim of this work is to obtain the Wigner rotation angle in the Schwarzschild spacetime via the Fermi-Walker transport of spinors. Then, from the WKB approach, we discuss a possible application of the Fermi-Walker transport of spinors in relativistic EPR correlations, where we show that the spins of the correlated particle undergo a precession in an analogous way to that obtained in Ref. \cite{tu2} via the application of successive infinitesimal Lorentz transformations. Furthermore, based on the WKB approach, we also discuss the degree of violation of the Bell inequality and the relativistic effects that stem from the geometry of the spacetime and the accelerated motion of the correlated particles in the nonrelativistic limit.

The structure of this paper is: in section II, we introduce the Fermi-Walker transport of spinors in a curved spacetime and in a quantum system; in section III, we calculate the Wigner rotation angle for a particle in a circular motion in the Schwarzschild spacetime via the Fermi-Walker transport of spinors; in section IV, we discuss the relativistic EPR correlations and the Bell inequality in the Schwarzschild spacetime via the Fermi-Walker transport of spinors. We also discuss the relativistic effects that stem from the geometry of the spacetime and the accelerated motion of the correlated particles is investigated in the nonrelativistic limit; in section V, we present our conclusions.

\section{The Fermi-Walker Transport}

In this section, we make a brief review of the Fermi-Walker transport of spinors in a curved spacetime. In a curved spacetime background, spinors are defined locally, where each spinor transforms according to the infinitesimal Lorentz transformations, that is, $\psi'\left(x\right)=D\left(\Lambda\left(x\right)\right)\,\psi\left(x\right)$, where $D\left(\Lambda\left(x\right)\right)$ corresponds to the spinor representation of the infinitesimal Lorentz group and $\Lambda\left(x\right)$ corresponds to the local Lorentz transformations \cite{weinberg}. Locally, the reference frame of the observers can be build via a noncoordinate basis $\hat{\Theta}^{a}=e^{a}_{\,\,\,\mu}\left(x\right)\,dx^{\mu}$, where the components $e^{a}_{\,\,\,\mu}\left(x\right)$ are called \textit{tetrads} and satisfy the relation: $g_{\mu\nu}\left(x\right)=e^{a}_{\,\,\,\mu}\left(x\right)\,e^{b}_{\,\,\,\nu}\left(x\right)\,\eta_{ab}$ \cite{weinberg,bd,naka}, where $\eta_{ab}=\mathrm{diag}(- + + +)$ is the Minkowski tensor. The inverse of the tetrads are defined as $dx^{\mu}=e^{\mu}_{\,\,\,a}\left(x\right)\,\hat{\Theta}^{a}$, and the relations $e^{a}_{\,\,\,\mu}\left(x\right)\,e^{\mu}_{\,\,\,b}\left(x\right)=\delta^{a}_{\,\,\,b}$ and $e^{\mu}_{\,\,\,a}\left(x\right)\,e^{a}_{\,\,\,\nu}\left(x\right)=\delta^{\mu}_{\,\,\,\nu}$ are satisfied.

Let us consider a spinor being transported from a point $x$ of the spacetime to another point $x'$, where there exists the action of external forces, but no torque. In this case, the law of transport is given by the Fermi-Walker transport \cite{misner,synge,steph}. Since the spinors are defined locally, then, the Fermi-Walker transport of a spinor is given by
\begin{eqnarray}
\frac{D}{d\tau}\,e^{a}_{\,\,\,\mu}\left(x\right)=-\frac{1}{c^{2}}\,\left[a_{\mu}\left(x\right)\,U^{\nu}\left(x\right)-U_{\mu}\left(x\right)\,a^{\nu}\left(x\right)\right]\,e^{a}_{\,\,\,\nu}\left(x\right),
\label{1.1}
\end{eqnarray} 
where $\frac{D}{d\tau}$ is the covariant derivative, $U^{\nu}\left(x\right)=\frac{dx^{\nu}}{d\tau}$ is the 4-velocity, $a^{\nu}\left(x\right)=U^{\mu}\left(x\right)\,\nabla_{\mu}\,U^{\nu}\left(x\right)$ is the 4-acceleration ($\nabla_{\mu}$ are the components of the covariant derivative) and $\tau$ is the proper time of a particle.

From the quantum mechanical point of view, the Fermi-Walker transport was introduced by Anandan \cite{anandan} by using the WKB approximation, where the wave function can be expanded as a superposition of locally plane waves. In this way, Anandan showed that if a particle is moving in an accelerated path, but there exists no torque, then, the wave packet is Fermi-Walker transported from a initial point $x$ to a final point $x'$. In this work, we modify the operator which determines this evolution of the wave packet in order to work with the spinorial algebra, therefore, the modified operator that gives rise to the Fermi-Walker transport of a spinor is given by 
\begin{eqnarray}
\hat{\Xi}=\hat{P}\exp\left(\frac{i}{4}\int\Omega_{\mu\,a\,b}\left(x\right)\,\,\Sigma^{ab}\,dx^{\mu}\right),
\label{1.2}
\end{eqnarray}
where $\hat{P}$ denotes the path ordering operator, $\Sigma^{ab}=\frac{i}{2}\left[\gamma^{a},\,\gamma^{b}\right]$ is the (spinorial) generator of the Lorentz transformations, $\gamma^{a}$ are the Dirac matrices defined in the Minkowski spacetime \cite{bd,greiner}. The $\gamma^{\mu}$ matrices are related to the $\gamma^{a}$ matrices via $\gamma^{\mu}=e^{\mu}_{\,\,\,a}\left(x\right)\gamma^{a}$ \cite{bd}. Besides, the object $\Omega_{\mu\,a\,b}\left(x\right)$ given in Eq. (\ref{1.2}) is defined as
\begin{eqnarray}
\Omega_{\mu\,\,\,b}^{\,\,\,a}\left(x\right)=\omega_{\mu\,\,\,b}^{\,\,\,a}\left(x\right)+\tau_{\mu\,\,\,b}^{\,\,a}\left(x\right).
\label{1.3}
\end{eqnarray}
The term $\omega_{\mu\,\,\,b}^{\,\,a}\left(x\right)$ is a connection 1-form which is also called as the spin connection \cite{naka,tu2}. This connection 1-form can be obtained by solving the the Maurer-Cartan structure equations in the absence of torsion $d\hat{\theta}^{a}+\omega^{a}_{\,\,\,b}\wedge\hat{\theta}^{b}=0$ \cite{naka} or via
\begin{eqnarray}
\omega_{\mu\,\,\,b}^{\,\,a}\left(x\right)=-e^{a}_{\,\,\nu}\left(x\right)\,\left[\partial_{\mu}e^{\nu}_{\,\,b}\left(x\right)+\Gamma^{\nu}_{\mu\beta}\,e^{\beta}_{\,\,b}\left(x\right)\right].
\label{1.4}
\end{eqnarray}
On the other hand, the second term given in Eq. (\ref{1.3}) was introduced by Anandan \cite{anandan} which arises from the action of external forces on the wave function, where it is given by
\begin{eqnarray}
\tau_{\mu\,\,\,b}^{\,\,a}\left(x\right)=\frac{a^{\nu}\left(x\right)}{c^{2}}\,\left[e_{\,\,\nu}^{a}\left(x\right)\,e_{b\mu}\left(x\right)-e_{\,\,\mu}^{a}\left(x\right)\,e_{b\nu}\left(x\right)\right].
\label{1.5}
\end{eqnarray}

As pointed out by Anandan \cite{anandan}, the connection 1-form given in Eq. (\ref{1.5}) can give rise to quantum effects such as the arising of geometric quantum phases associated with the Thomas precession.

\section{Wigner rotation in the Schwarzschild spacetime}

Our focus in this section is on the Wigner rotation in the Schwarzschild spacetime. We show that we can obtain the angle of the Wigner rotation by applying the Fermi-Walker transport on spinors. First of all, let us write the line element of the Schwarzschild spacetime: 
\begin{equation}
ds^{2}=-B\left(r\right)c^{2}dt^{2}+\frac{1}{B\left(r\right)}dr^{2}+r^{2}d\theta^{2}+r^{2}\sin^{2}\theta d\varphi^{2} 
 \label{3.1}
\end{equation}
where $B\left(r\right)=1-\frac{r_{s}}{r}$ and $r_{s}=\frac{2GM}{c^{2}}$. Now, in order that the spinors can be Fermi-Walker transported, we need to build a nonrotating frame where noninertial effects can be observed from the action of external forces without any influence of arbitrary rotations of the spatial axis of the local frame. This nonrotating frame is called the Fermi-Walker reference frame \cite{misner,synge,steph}. This reference frame can be built by taking $\hat{\Theta}^{0}=e^{0}_{\,\,\,t}\left(x\right)\,dt$, which means that the components of the noncoordinate basis form a rest frame for the observers at each instant, and the spatial components of the noncoordinate basis $\hat{\Theta}^{i}$ must be chosen in such a way that they do not rotate. From this conditions, we can write
\begin{eqnarray}
\hat{\Theta}^{0}=c\,\sqrt{B\left(r\right)}\,dt;\,\,\,\hat{\Theta}^{1}=\frac{1}{\sqrt{B\left(r\right)}}\,dr;\,\,\,\hat{\Theta}^{2}=r\,\sin\theta\,d\varphi;\,\,\,\hat{\Theta}^{3}=r\,d\theta.
\label{3.2}
\end{eqnarray}

By solving the Maurer-Cartan structure equations in the absence of torsion $d\hat{\theta}^{a}+\omega^{a}_{\,\,\,b}\wedge\hat{\theta}^{b}=0$ \cite{naka}, where $\omega^{a}_{\,\,\,b}=\omega_{\mu\,\,\,\,b}^{\,\,\,a}\left(x\right)\,dx^{\mu}$ is the connection 1-form, the operator $d$ corresponds to the exterior derivative and the symbol $\wedge$ means the wedge product, we obtain the following non-null components of the connection 1-form \cite{tu2}:
\begin{eqnarray}
\omega_{t\,\,\,\,1}^{\,\,\,0}\left(x\right)&=&-\omega_{t\,\,\,\,0}^{\,\,\,1}\left(x\right)=\frac{c}{2}\,\frac{r_{s}}{r^{2}};\nonumber\\
\omega_{\theta\,\,\,3}^{\,\,\,1}\left(x\right)&=&-\omega_{\theta\,\,\,1}^{\,\,\,3}\left(x\right)=-\sqrt{B\left(r\right)};\nonumber\\
[-2mm]\label{3.3}\\[-2mm]
\omega_{\varphi\,\,\,2}^{\,\,\,\,1}\left(x\right)&=&-\omega_{\varphi\,\,\,1}^{\,\,\,\,2}\left(x\right)=-\sqrt{B\left(r\right)}\sin\theta;\nonumber\\
\omega_{\varphi\,\,\,2}^{\,\,\,3}\left(x\right)&=&-\omega_{\varphi\,\,\,3}^{\,\,\,2}\left(x\right)=-\cos\theta.\nonumber
\end{eqnarray}

Henceforth, let us consider a circular motion with $r=\mathrm{const}$, where $r\,>\,r_{s}$, and $\theta=\pi/2$; thus, we can deal with the system in $\left(2+1\right)$ dimensions. The tangent vector associated with this motion is given by
\begin{eqnarray}
U^{t}\left(x\right)=\frac{1}{\sqrt{B(r)}}\,\cosh\xi,\,\,\,\,\,\,\,U^{\varphi}\left(x\right)=\frac{c}{r}\sinh\xi,
\label{3.4}
\end{eqnarray}
where $\tanh\xi=\frac{v}{c}$ and $v=\frac{r}{\sqrt{B\left(r\right)}}\,\frac{d\varphi}{dt}$. Thereby, there is one non-null component of the 4-acceleration which is given by
\begin{eqnarray}
a^{r}\left(x\right)=a=-c^{2}\,\sinh^{2}\xi\left\{\frac{B(r)}{r}-\frac{r_{s}}{2r^{2}}\,\coth^{2}\xi\right\}.
\label{3.5}
\end{eqnarray}

From the 4-acceleration (\ref{3.5}) and the local reference frame of the observers defined in Eq. (\ref{3.2}), we can calculate the connections 1-form $\tau^{\,\,\,a}_{\mu\,\,\,b}\left(x\right)$ \cite{bcf3}, then, we have (with respect to $\left(2+1\right)$ dimensions) 
\begin{eqnarray}
\tau^{\,\,\,0}_{t\,\,\,1}\left(x\right)&=&\tau^{\,\,\,1}_{t\,\,\,0}\left(x\right)=-\frac{a}{c};\nonumber\\
[-2mm]\label{3.6}\\[-2mm]
\tau^{\,\,\,1}_{\varphi\,\,\,2}\left(x\right)&=&-\tau^{\,\,\,2}_{\varphi\,\,\,1}\left(x\right)=\frac{a\,r}{c^{2}\sqrt{B(r)}}, \nonumber
\end{eqnarray}
therefore, the non-null components of $\Omega_{\mu\,\,\,b}^{\,\,\,a}\left(x\right)$ given in Eq. (\ref{1.3}) are
\begin{eqnarray}
\Omega^{\,\,\,0}_{t\,\,\,1}\left(x\right)&=&-\Omega^{\,\,\,1}_{t\,\,\,0}\left(x\right)=\frac{c}{2}\,\frac{r_{s}}{r^{2}}-\frac{a}{c};\nonumber\\
[-2mm]\label{3.7}\\[-2mm]
\Omega^{\,\,\,1}_{\varphi\,\,\,2}\left(x\right)&=&-\Omega^{\,\,\,2}_{\varphi\,\,\,1}\left(x\right)=\frac{a\,r}{c^{2}\sqrt{B(r)}}-\sqrt{B(r)}.\nonumber
\end{eqnarray}

Observe that, in this circular motion, the proper time of a particle is $\tau=\frac{r\,\Phi}{c\,\sinh\xi}$ \cite{tu2}, then, the operator (\ref{1.2}) that gives rise to the Fermi-Walker transport of a spinor, where the Dirac matrices $\gamma^{a}$ are given in $\left(2+1\right)$ dimensions, that is, $\gamma^{0}=\sigma^{3}$, $\gamma^{1}=i\sigma^{1}$ and $\gamma^{2}=i\sigma^{2}$ \cite{matrix,matrix2}. The matrices $\sigma^{k}$ are the Pauli matrices and satisfy the relation $\left(\sigma^{i}\,\sigma^{j}+\sigma^{j}\,\sigma^{i}\right)=2\,\delta^{ij}$. Thereby, the operator (\ref{1.2}) becomes \cite{bcf2}
\begin{eqnarray}
\hat{\Xi}=\exp\left(i\,\frac{\Gamma}{2}\right)=\sum_{n=0}^{\infty}\frac{1}{n!}\,\left(\frac{\Gamma}{2}\right)^{n}=\cos\frac{\alpha}{2}+i\,\frac{\Gamma}{\alpha}\,\sin\frac{\alpha}{2},
\label{3.8}
\end{eqnarray}
where $\Gamma$ and $\Gamma^{2}$ are matrices defined as
\begin{eqnarray}
\Gamma=\left(
\begin{array}{cc}
-\eta_{2} & -\eta_{1} \\
\eta_{1} & \eta_{2} \\
\end{array}\right);\,\,\,\,\,\,\,
\Gamma^{2}=\,\alpha^{2}\left(
\begin{array}{cc}
1 & 0 \\
0 & 1 \\
\end{array}\right)
\label{3.9}
\end{eqnarray}
whose parameters written in the matrices above are $\eta_{1}=\beta\,\Phi\,\sinh\xi\cosh\xi\,\sqrt{B\left(r\right)}$, $\eta_{2}=\beta\,\Phi\,\cosh^{2}\xi\,\sqrt{B\left(r\right)}$ and $\beta=1-\frac{r_{s}}{2\,r\,B\left(r\right)}$. Moreover, the parameter $\alpha$ established in Eqs. (\ref{3.8}) and (\ref{3.9}) is defined by  
\begin{eqnarray}
\alpha=\sqrt{\eta_{2}^{2}-\eta_{1}^{2}}=\Phi\,\sqrt{B\left(r\right)}\,\cosh\xi\left[1-\frac{r_{s}}{2\,r\,B\left(r\right)}\right].
\label{3.10}
\end{eqnarray}

The angle (\ref{3.10}) corresponds to the angle of the Wigner rotation in the Schwarzschild spacetime obtained via the Fermi-Walker transport of a spinor. Note that the Wigner rotation angle depends on the geometry of the spacetime, the parameter $\xi$ related to the accelerated motion of the particles and the position of the observers. This result agrees with Ref. \cite{tu2}, where the angle of the Wigner rotation is obtained through the application of infinitesimal Lorentz transformations.

Despite the connections 1-form given in Eqs. (\ref{3.6}) and (\ref{3.7}) are the same obtained in Ref. \cite{bcf3}, we should note that the discussion performed in Ref. \cite{bcf3} concerns the Fermi-Walker transport of vectors in two different scenarios of general relativity in the context of the classical physics. In this work, we deal with the Fermi-Walker transport of spinors in $\left(2+1\right)$ dimensions in the context of the quantum theory. Since classical paths cannot be established in a quantum system, we perform the WKB approximation in order to introduce a geometric approach yielded by the Fermi-Walker transport operator given in Eq. (\ref{3.8}) to study relativistic EPR correlations in a general relativity background. Thereby, we are able to apply the WKB approximation and show that these results can be extended to studies of relativistic EPR correlations in curved spacetime \cite{tu2,bcf,bcf2}. This is our goal in the next section.

\section{Relativistic EPR correlations in the Schwarzschild spacetime}

In this section, we wish to discuss the behaviour of a relativistic EPR correlation and the violation of the Bell inequality when the spinors are Fermi-Walker parallel transported in the Schwarzschild spacetime. In general relativity, the relativistic EPR quantum states can be defined as \cite{tu2}:
\begin{eqnarray}
\left|\psi^{\pm}\right\rangle&=&\frac{1}{\sqrt{2}}\left[\left|p_{+}^{a}\left(x\right),\,\uparrow;\,x\right\rangle\left|p^{a}_{-}\left(x\right),\,\downarrow;\,x\right\rangle\pm\left|p_{+}^{a}\left(x\right),\,\downarrow;\,x\right\rangle\left|p^{a}_{-}\left(x\right),\,\uparrow;\,x\right\rangle\right];\nonumber\\
[-2mm]\label{4.0}\\[-2mm]
\left|\phi^{\pm}\right\rangle&=&\frac{1}{\sqrt{2}}\left[\left|p_{+}^{a}\left(x\right),\,\uparrow;\,x\right\rangle\left|p^{a}_{-}\left(x\right),\,\uparrow;\,x\right\rangle\pm\left|p_{+}^{a}\left(x\right),\,\downarrow;\,x\right\rangle\left|p^{a}_{-}\left(x\right),\,\downarrow;\,x\right\rangle\right],\nonumber
\end{eqnarray}
where $p^{a}\left(x\right)$ is defined in the local reference frame of the observers, $x$ denotes the position of the observers and $\sigma=\uparrow,\,\downarrow$ denote the spins of the particles. Hence, by definition, each state above transforms under local Lorentz transformations in the spin-half representation at each local reference frame of the observers.

Let us consider an EPR source and two observers on the plane defined by $r=\mathrm{const}$, where $r\,>\,r_{s}$, and $\theta=\pi/2$, whose positions are given by the azimuthal angles $\varphi=0$ and $\varphi=\pm\Phi$, respectively. Thereby, we consider that an EPR pair of particles are emitted from the source in opposite directions in a circular motion, whose initial states are described by
\begin{eqnarray}
\left|\psi^{-}\right\rangle=\frac{1}{\sqrt{2}}\,\left\{\left|p_{+}^{a}\left(0\right),\uparrow;\,\varphi=0\right\rangle\left|p_{-}^{a}\left(0\right),\downarrow;\,\varphi=0\right\rangle-\left|p_{+}^{a}\left(0\right),\downarrow;\,\varphi=0\right\rangle\left|p_{-}^{a}\left(0\right),\uparrow;\,\varphi=0\right\rangle\right\},
\label{4.1}
\end{eqnarray}
where the 4-momentum of each particle in the local reference frame is given by $p_{\pm}^{a}\left(0\right)=m\,c\,\left(\cosh\xi,\,0,\,0,\,\pm\sinh\xi\right)$, and the spins are parallel to the 1-axis of the local reference frame as in Ref. \cite{tu2}. 

After the emission of the relativistic EPR pair of particles, we consider the WKB approximation in order to describe the Fermi-Walker transport of the quantum state (\ref{4.1}) from the initial point $\varphi=0$ to the final points $\varphi=\pm\Phi$ where the observers are placed. Note that the operator (\ref{3.8}) acts on each spin state of the particle, therefore we label $\hat{\Xi}_{\pm}=\cos\frac{\alpha}{2}\pm i\,\frac{\Gamma}{\alpha}\,\sin\frac{\alpha}{2}$; thus, the operator $\hat{\Xi}_{+}=\cos\frac{\alpha}{2}+i\,\frac{\Gamma}{\alpha}\,\sin\frac{\alpha}{2}$ acts on the spin states of the particle with momentum $p^{a}_{+}$, while the operator $\hat{\Xi}_{-}=\cos\frac{\alpha}{2}-i\,\frac{\Gamma}{\alpha}\,\sin\frac{\alpha}{2}$ acts on the spin states of the particle with momentum $p^{a}_{-}$. By considering the observers are in the rest frame of the particles given in Eq. (\ref{3.2}), then, after applying the Fermi-Walker transport (given by the operator (\ref{3.8})) on the quantum state (\ref{4.1}), we obtain the following quantum state in the points $\varphi=\pm\Phi$ where the observers are placed:
\begin{eqnarray}
\left|\zeta\right\rangle=\cos\alpha\,\left|\psi^{-}\right\rangle-i\,\cosh\xi\,\sin\alpha\,\left|\psi^{+}\right\rangle+i\,\sinh\xi\,\sin\alpha\,\left|\phi^{+}\right\rangle.
\label{4.2}
\end{eqnarray}

Due to the Fermi-Walker transport operator (\ref{3.8}) acts on the spinors, that is, on the spin states, thus, the quantum state of the correlated particles obtained in Eq. (\ref{4.2}) is analogous to the quantum states obtained in Ref. \cite{tu2} via infinitesimal Lorentz transformations (applied at each point of the spacetime from $\varphi=0$ to $\varphi=\pm\Phi$) in the sense that the spins of the relativistic correlated particles in the initial EPR correlation undergo a precession that depends on the angle $\alpha$, and and the 4-momentum of the particles remains unchanged in the local reference frame of the observers \cite{anandan}. However, we can see that the final state of the relativistic correlated particles is given by the superposition of the states $\left|\psi^{-}\right\rangle$, $\left|\psi^{+}\right\rangle$ and $\left|\phi^{+}\right\rangle$ and depends on the angle of the Wigner rotation $\alpha$ and the parameter $\xi$, which differs from the state obtained in Ref. \cite{tu2}. Moreover, no spurious effects from arbitrary rotations of the local axis (as mentioned in Ref. [14]) exist, since the Fermi-Walker reference frame is built in such a way that the local axis do not rotate.

As pointed out in Refs. \cite{tu2,bcf}, the states of the relativistic correlated particles after undergoing the Wigner rotation suggests that the initial spin anticorrelations given in Eq. (\ref{4.1}) are broken if the observers measure the spin in the direction of the 1-axis. Actually, the quantum state (\ref{4.2}) shows that the direction of the spin measurements in which the observers can make at the points $\varphi=\pm\Phi$ must be rotated about the 3-axis of the local reference frame of observers (given in Eq. (\ref{3.2})) through the angles $\mp\alpha$, respectively. This means that, by knowing the relativistic effects from the acceleration of the particles and the geometry of the spacetime, we can recover the initial spin anticorrelations given in Eq. (\ref{4.1}) by rotating the spin axis of measurement through the angles $\mp\alpha$ about the 3-axis of the local reference frame established in Eq. (\ref{3.2}) at $\varphi=\pm\Phi$.

Henceforth, let us analyse the violation of the Bell inequality in this system, where the particles are moving in a circular motion on the plane defined by $r=\mathrm{const}$, where $r\,>\,r_{s}$, and $\theta=\pi/2$. Suppose the first observer is placed in $\varphi=+\Phi$ and measure the component of the spin through the observables $\hat{a}$ and $\hat{a}'$, while the second observer is placed in $\varphi=-\Phi$ and measure the component of the spin through the observables $\hat{b}$ and $\hat{b}'$, where these operators are defined as
\begin{eqnarray}
\hat{a}=\frac{\sigma^{1}+\sigma^{3}}{\sqrt{2}};\,\,\,\hat{a}'=\frac{\sigma^{3}-\sigma^{1}}{\sqrt{2}};\,\,\,\hat{b}=\sigma^{3};\,\,\,\,\hat{b}'=\sigma^{1}.
\label{4.3}
\end{eqnarray} 

It is well-known that the maximum violation of the Bell inequality can be given by the following expression \cite{nc}:
\begin{eqnarray}
\left|\left\langle \hat{a}\,\hat{b}\right\rangle+\left\langle \hat{a}'\,\hat{b}\right\rangle+\left\langle \hat{a}\,\hat{b}'\right\rangle-\left\langle \hat{a}'\,\hat{b}'\right\rangle\right|\leq2\,\sqrt{2}.
\label{4.4}
\end{eqnarray}

Thereby, let us analyse the relativistic effects which arises from gravity and the accelerated motion of the particles on the Bell inequality. By taking the quantum state given in Eq. (\ref{4.2}), we obtain
\begin{eqnarray}
\left|\left\langle \hat{a}\,\hat{b}\right\rangle+\left\langle \hat{a}'\,\hat{b}\right\rangle+\left\langle \hat{a}\,\hat{b}'\right\rangle-\left\langle \hat{a}'\,\hat{b}'\right\rangle\right|=2\,\sqrt{2}\,\left|\cosh^{2}\xi\,\sin^{2}\alpha-1\right|.
\label{4.5}
\end{eqnarray}

We can see in Eq. (\ref{4.5}) that the Bell inequality depends on the angle of the Wigner rotation $\alpha$ given in Eq. (\ref{3.10}) and the parameter $\xi$. Observe that the degree of violation of the Bell inequality obtained from the quantum state of the correlated particles (\ref{4.2}), given by the Fermi-Walker transport of spinors, is different with respect to that obtained in Ref. \cite{tu2}. This dependence on the angle $\alpha$ and the parameter $\xi$ means that the degree of violation of Bell inequality decreases by making the spin measurements on the 1-axis of the local reference frame of the observers at $\varphi=\pm\Phi$. This result stems from the relativistic effects of the geometry of the spacetime and the accelerated motion of the correlated particles in the plane defined by $\theta=\pi/2$ and $r=\mathrm{const}$, where $r\,>\,r_{s}$. However, as we have discussed previously, by rotating the the spin axes of measurements through the angles $\mp\alpha$ about the 3-axis of the local reference frame of the observers (given in Eq. (\ref{3.2})) at $\varphi=\pm\Phi$, we can recover both the initial spin anticorrelations and the maximum violation of the Bell inequality given in Eq. (\ref{4.4}).

Finally, let us consider $\frac{v}{c}\ll1$, which corresponds to the nonrelativistic limit. By considering $\frac{v}{c}\ll1$, then, the quantum state (\ref{4.2}) becomes
\begin{eqnarray}
\left|\bar{\zeta}\right\rangle=\cos\bar{\alpha}\,\left|\psi^{-}\right\rangle-i\,\sin\bar{\alpha}\,\left|\psi^{+}\right\rangle,
\label{4.6}
\end{eqnarray}
where the angle $\bar{\alpha}$ corresponds to the nonrelativistic limit of the angle of the Wigner rotation, which is given by
\begin{eqnarray}
\bar{\alpha}\approx\Phi\,\left[1-\frac{r_{s}}{2\,r\,B\left(r\right)}\right]\,\sqrt{B\left(r\right)}\left(1+\frac{v^{2}}{2c^{2}}\right).
\label{4.7}
\end{eqnarray}
Moreover, the result obtained in Eq. (\ref{4.5}) which corresponds to the Bell inequality becomes
\begin{eqnarray}
\left|\left\langle \hat{a}\,\hat{b}\right\rangle+\left\langle \hat{a}'\,\hat{b}\right\rangle+\left\langle \hat{a}\,\hat{b}'\right\rangle-\left\langle \hat{a}'\,\hat{b}'\right\rangle\right|=2\,\sqrt{2}\,\cos^{2}\bar{\alpha}.
\label{4.8}
\end{eqnarray}

We can see from Eqs. (\ref{4.6}), (\ref{4.7}) and (\ref{4.8}) that relativistic effects from the geometry of the spacetime and the accelerated motion of the correlated particles exist in the nonrelativistic limit. We can also see a different state of the correlated particles (given by Eq. (\ref{4.6})) in the nonrelativistic limit, despite this state of the nonrelativistic correlated particles is rotated in an analogous way to the relativistic case discussed previously, however, this rotation depends only on the angle of rotation is given in Eq. (\ref{4.7}). In addition, the degree of violation of the Bell inequality is changed with respect to the relativistic case given in Eq. (\ref{4.5}). In Eq. (\ref{4.8}), we have that the degree of violation of the Bell inequality decreases due to the relativistic effects from the geometry of the spacetime and the accelerated motion of the correlated particles, and it is determined by the angle of rotation is given in Eq. (\ref{4.7}). We can also observe that the expression of the degree of violation of the Bell inequality given in Eq. (\ref{4.8}) in the nonrelativistic limit differs from that discussed in Ref. \cite{tu2}. However, the initial spin anticorrelations and the maximum violation of the Bell inequality are recovered by rotating the the spin axes of measurements through the angles $\mp\bar{\alpha}$ about the 3-axis of the local reference frame of the observers (given in Eq. (\ref{3.2})) at $\varphi=\pm\Phi$.

\section{conclusions}

We have discussed the Fermi-Walker transport of spinors in the Schwarzschild spacetime and shown, by considering a particle moving in a circular path on the plane defined by $r=\mathrm{const}$ (where $r\,>\,r_{s}$) and $\theta=\pi/2$, that the angle of the Wigner rotation can be obtained, where the Wigner rotation angle depends on the geometry of the spacetime and the parameter $\xi$ related to the accelerated motion of the particles and the position of the observers. This results agrees with the results of Ref. \cite{tu2} obtained by applying infinitesimal Lorentz transformations. 

We have also discussed a possible application of this study in relativistic EPR correlations. By considering the WKB approximation, we have shown that when the spinors of an EPR pair of particle are Fermi-Walker transported, then, the spins of the correlated particles undergo a precession which depends on the angle of the Wigner rotation and the 4-momentum of each particle remains unchanged in the local reference frame of the observers. This precession of spins suggests that the initial spin anticorrelations are broken if each observer make the spin measurement in the direction of the 1-axis. However, the final quantum states which describes the spin precession indicates that the direction of the spin measurements at the final points $\varphi=\pm\Phi$ must be rotated about the 3-axis of the local reference frame of the observers through the angles $\mp\alpha$ (Wigner rotation angle), respectively. Therefore, by knowing the relativistic effects from the acceleration of the particles and the geometry of the spacetime, we can recover the initial spin anticorrelations by correcting the direction of the spin axis in terms of the Wigner rotation angle. 

We have extended our discussion to the Bell inequality by showing that the Bell inequality depends on the angle of the Wigner rotation, whose meaning is that the degree of violation of Bell inequality decreases by making the spin measurements on the 1-axis of the local reference frame of the observers at $\varphi=\pm\Phi$. This decrease in the Bell inequality stems from the accelerated motion of the correlated particles and the relativistic effects of the geometry of the spacetime. On the other hand, we have seen that by rotating the spin axes of measurements through the angles $\mp\alpha$ about the 3-axis of the local reference frames of the observers at $\varphi=\pm\Phi$, we can recover the maximum violation of the Bell inequality.

Finally, we have discussed a particular case corresponding the nonrelativistic limit given by $\frac{v}{c}\ll1$, and observed that the relativistic effects from the geometry of the spacetime and the accelerated motion of the correlated particles exist in the nonrelativistic limit. The Fermi-Walker transport of spinors gives rise to the precession of the spins of the correlated particles and the degree of violation of the Bell inequality decreases in an analogous way to the relativistic case. Again, we have that the initial spin anticorrelations and the maximum violation of the Bell inequality can be recovered by rotating the the spin axes of measurements through the appropriated angles.

\acknowledgements

The authors would like to thank CNPq (Conselho Nacional de Desenvolvimento Cient\'ifico e Tecnol\'ogico - Brazil) for financial support.

\end{document}